# Attack Identification and Correction for PMU GPS Spoofing in Unbalanced Distribution Systems

Ying Zhang, *Student Member, IEEE*, Jianhui Wang, *Senior Member, IEEE*, Jianzhe Liu, *Member, IEEE*

*Abstract*—Due to the vulnerability of civilian global positioning system (GPS) signals, the accuracy of phasor measurement units (PMUs) can be greatly compromised by GPS spoofing attacks (GSAs), which introduce phase shifts into true phase angle measurements. Focusing on simultaneous GSAs for multiple PMU locations, this paper proposes a novel identification and correction algorithm in distribution systems. A sensitivity analysis of state estimation residuals on a single GSA phase angle is firstly implemented. An identification algorithm using a probing technique is proposed to determine the locations of spoofed PMUs and the ranges of GSA phase shifts. Based on the identification results, these GSA phase shifts are determined via an estimation algorithm that minimizes the mismatch between measurements and system states. Further, with the attacked PMU data corrected, the system states are recovered. Simulations in unbalanced IEEE 34-bus and 123-bus distribution systems demonstrates the efficiency and accuracy of the proposed method.

*Index Terms*—State estimation, phasor measurement units, multiple GPS spoofing attacks, unbalanced distribution systems, attack identification and correction.

## I. Introduction

CYBER security is a critical issue for the reliability of power networks [1]. The 2015 Ukraine blackout was a first-of-its-kind cyber incident that resulted in nationwide power outages covering 225,000 customers for several hours. Motivated by this event, increasing amounts of research are being carried out on this issue, especially the cyber-attacks against phasor measurement units (PMUs) [2]. Also, installation of PMUs increasingly makes dramatic changes to the landscape of power grid operation and enables more-accurate state estimation [3].

Wide use of advanced information and communication technology creates opportunities for global positioning system (GPS) spoofing attacks (GSAs) against PMUs that rely on civilian GPS timing signals. GSAs compromise the synchronization of measuring devices by introducing forged GPS signals [4]. The approaches that GSAs manipulate the GPS signals to introduce phase offsets into PMU data are demonstrated in [5] and [6], where the experimental and theoretical feasibility were verified. Moreover, the North American Electric Reliability Corporation reported a real-world GSA in 2012 [7]. As a result, GSAs create the mismatch between the measured and true phase angles of synchronous data; these negate the effectiveness of PMU measurements.

Recent studies on GSAs mainly focus on two issues: (1) the impacts and analysis of GSAs (*e.g.*, [5], [8], and [9]), and (2) detection and correction for GSAs in transmission systems (*e.g.*, [10]–[13]). For example, a detection mechanism for multiple spoofing attacks is proposed from a physical perspective in [10]. This mechanism requires another commercial GPS receiver to be installed close to the existing one in a PMU. Most bad data detection techniques directly remove bad data once detected, provided that there is a certain level of measurement redundancy, and that these bad data individually occur. However, removing the spoofed data under multiple GSAs may cause the system unobservable, as a single GSA can not only compromise the voltage phasor at a bus but also the current phasors of several branches connected to the attacked bus. Hence, the authors of [11] presented a correction algorithm for a single GSA in transmission systems with only PMU measurements, and this method performs the generalized likelihood ratio tests presented in [12] for correcting the GSA on each candidate PMU. However, the number of PMUs in distribution systems is insufficient to make the entire system observable due to technical and economic limits. Moreover, this algorithm cannot achieve correction for multiple GSAs. Risbud *et al.* [13] proposed an alternating minimization algorithm for these unknown spoofed PMU locations and phase angels for the attack reconstruction of multiple GSAs. Beyond these mentioned works focusing on the countermeasures to GPS spoofing, other works on the attack identification for PMUs in transmission systems include [14] and [15], which have specific requirements for the amount of PMU data. For instance, [14] requires continuous PMU measurements across time for dynamic state estimation, and [15] realizes the PMU data recovery specifically focusing on a transmission line equipped with two PMUs at both ends.

As reviewed in [16], distribution systems are also vulnerable to cyber-attacks due to their direct connections to customer loads and emerging distributed generators (DGs). However, the existing state estimation-based algorithms in transmission systems cannot be trivially applied to unbalanced distribution systems with high r/x ratios [3]. Moreover, in contrast to extensive works regarding the cyber-attacks against state estimation in transmission systems, in the literature, only a handful of studies (*e.g.*, [17] – [21]) address the cyber-attack issues in distribution networks. For example, the impact of cyber-attacks on voltage regulation is considered in [17] with photovoltaic (PV) devices connected. Recently, the authors of [20] proposed a false data injection attack mechanism against distribution system state estimation (DSSE) in balanced systems from an attacker's point of view. Then, they extended

Y. Zhang and J. Wang are with Electrical and Computer Engineering Department, Southern Methodist University, Dallas, TX 75205, USA (email: yzhang1@smu.edu; jianhui@smu.edu).

J. Liu is with Argonne National Laboratory, Energy Systems Division, Lemont, IL 60439, USA (email: jianzhe.liu@anl.gov).

[20] to multiphase and unbalanced distribution systems in [21], since the multiphase and unbalanced natures are typical features of practical systems and demand further research. However, these existing studies focus on the construction of cyber-attacks and thus do not yet address the correction of GSAs against PMU data, and the related research in transmission systems such as [11] – [14] cannot be applied to unbalanced distribution systems. Hence, with more distribution-level PMUs installed in systems, the identification and correction of GSAs in unbalanced distribution systems call for solutions.

This paper proposes a novel algorithm to identify and correct the corrupted PMU data under multiple GSAs in unbalanced distribution systems. Compared with a single GSA, multiple GSAs refer to those GSAs that are simultaneously launched on multiple PMU locations [13]. It is observed that the meters located in substations with higher cyber-security could be more challenging to access by attackers [16]. The proposed algorithm of GSAs is based on this observation, which is supported in Section III-A. This paper presents an identification method using a probing technique, and probing is defined as the technique of perturbing measurements to find the locations of spoofed PMUs and the ranges of GSA phase shifts. The idea of probing is widely adopted in algorithmic applications to power system operation, *e.g.*, topology identification of distribution networks in [22]. Based on the identification method, these GSA phase shifts are obtained by minimizing the mismatch between all the measurements and state variables. Further, with the attacked PMU data corrected, the system states are recovered.

The contributions of this paper are summarized as follows:
- The proposed method enables the identification and correction of multiple GSAs by using DSSE algorithms with PMU measurements and smart meter data. The hybrid measurement deployment becomes pervasive for current distribution networks, which is not fully accounted for in the existing literature. Moreover, the algorithm is not limited by the number of spoofed PMU locations.
- The proposed algorithm is hierarchical and more efficient. Specifically, the proposed identification method first determines the locations of attacked PMUs and the magnitude ranges of GSA phase shifts, then accelerates the subsequent correction process.
- This algorithm is applied to multiphase and unbalanced distribution systems and considers the impacts of DG penetration.
- The proposed identification method has no requirement for additional hardware and could be transplanted before other correction algorithms.

II. PRELIMINARY

*A. DSSE Integrating PMU Data*

In a classical state estimator, the relationship between measurements and state variables is expressed as:

$$z = h(x) + e \quad (1)$$

where $z$ is a measurement vector, and $z \in \mathbb{R}^{m \times 1}$; $h(x)$ is the measurement function about the state vector $x$, and $x \in \mathbb{R}^{n \times 1}$; the measurement noise vector $e \sim N(0, R)$ follows Gaussian distributions with the covariance matrix $R = diag[\sigma_1^2, \sigma_2^2, \ldots, \sigma_m^2]$, where $\sigma_j^2$ represents the variance of the $j$th measurement noise, $j = 1, \ldots, m$, and the measurement weight matrix is defined as $W = R^{-1}$.

A weighted least square (WLS) criterion is used to minimize the sum of weighted measurement residuals (WMRs), $J$:

$$J = [z - h(x)]^T W [z - h(x)] = r^T W r \quad (2)$$

where $r = z - h(x)$ is the measurement residual vector, and $[\cdot]^T$ denotes the transpose of vectors or matrices.

The estimated state vector is obtained by Newton's method until $\Delta x$ at iteration $t$ is less than a pre-set tolerance:

$$\nabla J = 0 \quad (3)$$

$$\Delta x = (H(x^t)^T W H(x^t))^{-1} H(x^t)^T W [z - h(x^t)] \quad (4)$$

$$x^{t+1} = x^t + \Delta x \quad (5)$$

where $H(x^t)$ is the Jacobian matrix and $H(x^t) = \partial h(x^t)/\partial x^t$.

Developed from the classical estimator, an efficient DSSE method integrating PMU and smart meter data proposed in [23] is used in this paper. In this DSSE algorithm, the voltage at the substation bus and branch currents are chosen as state variables. A general three-phase distribution line model is shown in Fig.1, and the state variables of the three-phase system are expressed in rectangular coordinates:

$$x = [v_{slack,r}^a, v_{slack,x}^a, \ldots, v_{slack,x}^c, i_{1r}^a, i_{1x}^a, \ldots, i_{Nx}^c]^T \quad (6)$$

where $v_{slack,r}^\varphi$ and $v_{slack,x}^\varphi$ are the real and imaginary parts of the three-phase substation voltage, and $i_{pr}^\varphi$ and $i_{px}^\varphi$ are the real and imaginary parts of the three-phase current at branch $p$, $p = 1, \ldots, N$; the superscripts $\varphi \in \{a, b, c\}$ denote phase indices.

In this estimator, smart meters provide the power data at load/DG buses, while PMUs record the measurements of the magnitudes and phase angles of voltages and currents. Moreover, the relationship between the measurement functions and these measurements are listed below:

$$\begin{cases} h_{V_{kr}}(x) = z_{V_{kr}}, & k \in \psi_V \\ h_{V_{kx}}(x) = z_{V_{kx}}, & k \in \psi_V \end{cases} \quad (7)$$

$$\begin{cases} h_{I_{pr}}(x) = z_{I_{pr}}, & p \in \psi_I \\ h_{I_{px}}(x) = z_{I_{px}}, & p \in \psi_I \end{cases} \quad (8)$$

$$\begin{cases} h_{P_k}(x) = z_{P_k}, & k \in \psi_S \\ h_{Q_k}(x) = z_{Q_k}, & k \in \psi_S \end{cases} \quad (9)$$

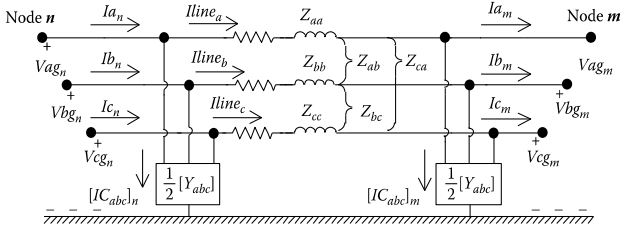

Fig.1. Three-phase line model in distribution systems

where these measurements can be expressed as 1) the real and imaginary parts of voltage $z_{V_{kr}}$ and $z_{V_{kx}}$, 2) the real and imaginary parts of current $z_{I_{pr}}$ and $z_{I_{px}}$, and 3) the active and reactive powers $z_{P_k}$ and $z_{Q_k}$; $h_{V_{kr}}(x)$, $h_{V_{kx}}(x)$, $h_{I_{pr}}(x)$, $h_{I_{px}}(x)$, $h_{P_k}(x)$, and $h_{Q_k}(x)$ are the corresponding measurement functions; $k$ is the index of buses, and $p$ is the index of branches; $\psi_V$ and $\psi_S$ denote the sets of the buses with the voltage and power measurements, and $\psi_I$ is the set of the branches that the current measurements are located at.

Moreover, for $k \in \psi_S$, the measurements at load/DG nodes are converted into equivalent currents by

$$z^{eq}_{I_{kr}} + jz^{eq}_{I_{kx}} = \left[\frac{z_{P_k} + jz_{Q_k}}{V_k}\right]^* \quad (10)$$

where $z^{eq}_{I_{kr}}$ and $z^{eq}_{I_{kx}}$ are the real and imaginary parts of the equivalent injection current at bus $k$, and $V_k$ is the voltage phasor at this bus; $[\cdot]^*$ denotes the complex conjugate.

The measurement vector $z$ is expressed as

$$z = [z_V \ z_I \ z_S]^T \quad (11)$$

where $z_V$ denotes the vector composed of the real and imaginary parts of three-phase voltages, $z_V = \begin{bmatrix} z_{V_{kr}} \\ z_{V_{kx}} \end{bmatrix}_{k \in \psi_V}$, $z_{V_{kr}} = [z^a_{V_{kr}}, z^b_{V_{kr}}, z^c_{V_{kr}}]^T$, and $z_{V_{kx}} = [z^a_{V_{kx}}, z^b_{V_{kx}}, z^c_{V_{kx}}]^T$; $z_I$ and $z_S$ are the vectors of the real and imaginary parts of current and equivalent current measurements, $z_I = \begin{bmatrix} z_{I_{pr}} \\ z_{I_{px}} \end{bmatrix}_{p \in \psi_I}$, and $z_S = \begin{bmatrix} z^{eq}_{I_{kr}} \\ z^{eq}_{I_{kx}} \end{bmatrix}_{k \in \psi_S}$.

The DSSE model with the use of PMU data has constant Jacobian elements, i.e., $H(x) = H$, briefly listed below.

1) PMU voltage measurements

For the PMU voltage at node $k$, $k \in \psi_V$, the measurement function is expressed as

$$h_{V_{kr}} + jh_{V_{kx}} = (v_{slack,r} + jv_{slack,x}) - \sum_{p \in \Im_k}(i_{pr} + ji_{px})Z_p \quad (12)$$

where $\Im_k$ is a set of line segments from the slack node to node $k$, $p \in \Im_k$, and $v_{slack,r}$, $v_{slack,x}$, $i_{pr}$, and $i_{px}$ are three-phase state vectors, e.g., $v_{slack,r} = [v^a_{slack,r} \ v^b_{slack,r} \ v^c_{slack,r}]^T$; $Z_p$ is the 3×3 impedance matrix at branch $p$, expressed as:

$$Z_p = \begin{bmatrix} Z_{aa} & Z_{ab} & Z_{ac} \\ Z_{ba} & Z_{bb} & Z_{bc} \\ Z_{ca} & Z_{cb} & Z_{cc} \end{bmatrix}$$

where the diagonal and off-diagonal elements such as $Z_{aa}$ and $Z_{ab}$ represent the self-impedances and mutual impedances between two phases, illustrated as Fig.1. See the constant Jacobian elements of (12) in [24].

2) PMU current measurements

For the PMU current at branch $p$, $p \in \psi_I$, $h_{I_{pr}} + jh_{I_{px}} = i_{pr} + ji_{px}$, and the Jacobian elements on phase $\varphi$ are shown as:

$$\frac{\partial h^\varphi_{I_{pr}}}{\partial i^\gamma_{lr}} = \begin{cases} 1, & \text{when } p = l \text{ and } \varphi = \gamma \\ 0, & \text{elsewhere} \end{cases}$$

$$\frac{\partial h^\varphi_{I_{px}}}{\partial i^\gamma_{lx}} = \begin{cases} 1, & \text{when } p = l \text{ and } \varphi = \gamma \\ 0, & \text{elsewhere} \end{cases}$$

where $l$ is the index of all branches, and the phase index $\gamma \in \{a, b, c\}$.

3) Power measurements from smart meters

For the powers at node $k$, $k \in \psi_S$, the Jacobian elements only has nonzero values of 1 and −1, the measurement function holds:

$$h_{I_{kr}} + jh_{I_{kx}} = i_{in,r} + ji_{in,x} - \sum(i_{out,r} + ji_{out,x}) \quad (13)$$

where $i_{in,r}$ and $i_{in,x}$ as state variables denote the real and imaginary parts of the input current at node $k$, and $i_{out,r}$ and $i_{out,x}$ denote the output current. For simplicity, the phase indices are suppressed here.

The DSSE process with this constant Jacobian matrix is implemented in the following steps [24]:

1) *Backward Sweep*: Get the initial branch currents by a backward approach. An initial voltage at each node is set as a substation voltage $V_{slack}$, and (10) is modified to calculate current injections as:

$$z^{eq}_{I_{kr}} + jz^{eq}_{I_{kx}} = \left[\frac{z_{P_k} + jz_{Q_k}}{V_{slack}}\right]^* \quad (14)$$

where these injections are used to obtain the initial branch currents, $x^0$.

2) *Forward Sweep*: The latest branch currents and the substation voltage are used to calculate the initial nodal voltages.

3) In iteration $t$, calculate $h(x^t)$, and then update system state variables by $\Delta x = (H^T W H)^{-1} H^T W [z - h(x^t)]$.

4) Update the branch currents by $x^{t+1} = x^t + \Delta x$, then calculate the new nodal voltages by the forward sweep.

5) If $\Delta x$ is less than a pre-set tolerance, stop the iterative process. Otherwise, use these updated voltages to calculate the equivalent currents by (10), then go to step 3).

### B. Multiple GSA Model

Multiple GSAs introduce clock offset errors to PMUs, and equivalently result in the phase shifts of these PMU data [8]. Without loss of generality, suppose that a GSA on PMU $i$ that is installed at bus $k$ and at branch $p$ introduces one clock offset error $\Delta t_i$, $k \in \psi_V$ and $p \in \psi_I$, where $i=1, \ldots, N_{pmu}$, and $N_{pmu}$ is the amount of the installed PMUs. The phase shift $\theta_i^{spf}$ corresponding to the clock offset error is denoted by $\theta_i^{spf} = 2\pi f \Delta t_i [rad]$. Hence, the GSA on a PMU affects the three-phase voltage and currents measured by the attacked PMU with the same phase shift [13].

Under no attacks, the PMU voltage at node $k$ measured by PMU $i$ are expressed as

$$z_{V,k} = [\ |V_k^a|\cos\theta_{V,k}^a, |V_k^a|\sin\theta_{V,k}^a, \ldots, |V_k^c|\sin\theta_{V,k}^c\ ]^T \quad (15)$$

where $z_{V,k} \in \mathbb{R}^{6\times 1}$, and $|V_k^\varphi|$ and $\theta_{V,k}^\varphi$ denote the $\varphi$-phase voltage magnitude and voltage phase angle at node $k$.

Under GSAs, the phase angles in (15) are shifted by $\theta_i^{spf}$, and the spoofed voltage vector $z_{V,k}^{spf}$ is expressed as

$$\begin{aligned} z_{V,k}^{spf} &= A_k z_{V,k} \\ &= [\ |V_k^a|\cos\left(\theta_{V,k}^a + \theta_i^{spf}\right), \ldots, |V_k^c|\sin\left(\theta_{V,k}^c + \theta_i^{spf}\right)\ ]^T \end{aligned} \quad (16)$$

where $z_{V,k}^{spf} \in \mathbb{R}^{6\times 1}$, $A_k = \text{diag}[A_i^a, A_i^b, A_i^c]$, and $A_i^\varphi = \begin{bmatrix} \cos\theta_i^{spf} & -\sin\theta_i^{spf} \\ \sin\theta_i^{spf} & \cos\theta_i^{spf} \end{bmatrix}$.

For the current at branch $p$, $p \in \psi_I$, $z_{I,p}^{spf} = A_p z_{I,p}$, where $z_{I,p}$ and $z_{I,p}^{spf}$ denote the normal and attacked current measurements, respectively, and $A_p = \text{diag}[A_i^a, A_i^b, A_i^c]$.

Considering GSAs on multiple PMUs, a spoofed measurement vector as an attack model is expressed below:

$$z_{spf} = \begin{bmatrix} A_V z_V \\ A_I z_I \\ z_S \end{bmatrix} = \begin{bmatrix} A_V & 0 & 0 \\ 0 & A_I & 0 \\ 0 & 0 & I \end{bmatrix} \begin{bmatrix} z_V \\ z_I \\ z_S \end{bmatrix} = Az \quad (17)$$

where $A_V$ and $A_I$ are the block matrices of the PMU voltage and current measurements affected by GSAs, and the diagonal block for smart meter data are an identity matrix, $I$. The diagonal blocks $A_V$ and $A_I$ are expressed as

$$A_V = \text{diag}[A_1, \ldots, A_k, \ldots] \qquad k \in \psi_V \quad (18.a)$$

$$A_I = \text{diag}[A_1, \ldots, A_p, \ldots] \qquad p \in \psi_I \quad (18.b)$$

Under multiple GSAs, the original measurement vector $z$ is unexpectedly replaced by $z_{spf}$ in (17). The phase shifts in $z_{spf}$ are unknown and arbitrarily vary in $[0, 2\pi]$ or $[-\pi, \pi]$. Without the knowledge of $A_V$ and $A_I$, or equivalently the locations and sizes of $\theta_i^{spf}$, the states cannot be accurately estimated by the only available measurement vector $z_{spf}$ in the attacked system. There is only one variable in the correction problem of a single GSA, while there are $N$ variables in the correction problem of multiple ($N$) GSAs, and $N$ is unknown. Moreover, it is difficult for system operators to know if the attack is a single GSA or multiple GSAs *a priori*.

## III. PROPOSED ALGORITHM

A set of synergistic mechanism is proposed to solve the correction problem of multiple GSAs (coordinated[1] or uncoordinated) in distribution systems which includes single GSAs as a special case. This mechanism is conducted hierarchically, including the identification and correction steps:

1) The identification method for multiple GSAs first enables the locations of the attacked PMUs, *i.e.*, the locations of those $\theta_i^{spf}$ that are not zero. Further, this algorithm determines narrower intervals than $[-\pi, \pi]$ that these non-zero $\theta_i^{spf}$ lie in.

2) Finally, an optimization model is formulated to obtain the values of these $\theta_i^{spf}$.

### A. Assumption

The proposed algorithm is based on the assumption that only the PMU in a substation is secure, *i.e.*, free of GSAs. Consequently, $A_1$ is an identity matrix with $\theta_1^{spf} = 0$ in (18).

Secure measurements protected from attacks could be obtained via a combination of encryption, authentication tags, continuous monitoring, and other tactics [26]. Moreover, substation measurements generally have higher cyber-security due to additional measurement protection schemes and data authentication. For instance, [27] explores currently available security solutions and studies their applicability to substations, while [28] as an industrial standard is published in 2007, and describes data security mechanisms to be deployed in current substations. Moreover, the security protocols against cyber-attacks in distribution automation systems are proposed in [29].

Besides, the assumption of secure measurements has been used in existing algorithmic studies (*e.g.*, [26] and [30]). For instance, [30] proposes a robust detection method for bad data injection with a number of secure PMU measurements. Also, the assumption in the proposed method that there is only one secure PMU in a substation is modest and practical.

### B. Linear DSSE Algorithm

Owing to the constant Jacobian matrix in Section II-A, the general formula (4) is updated with $h(x) = Hx$. Also, to avoid the iterative estimation process in Section II-A, the nodal voltages in (10) are fixed as $V_{slack}$ to obtain the equivalent current measurements, *i.e.*, (14). Then, a closed-form solution in the DSSE method is estimated as [21]:

$$\hat{x} = (H^T W H)^{-1} H^T W z = G^{-1} H^T W z \quad (19)$$

$$r = z - H\hat{x} = Kz \quad (20)$$

where $G$ is the gain matrix of this estimator, and $G = H^T W H$; $K = I - H G^{-1} H^T W$, and $K$ is defined as a residual sensitivity matrix and remains invariable due to known network structures and parameters.

---
[1]Coordinated GSAs are a type of coordinated or unobservable attacks, which are defined as the attacks that are well designed and coordinated to enable passing the bad measurement detection by evaluating the measurement residual $J(\hat{x})$ [25].

The closed-form DSSE algorithm is adopted in the identification method of Section III-C due to its high computational efficiency.

### C. Identification of Multiple GSAs

In this subsection, the probing technique is performed on each PMU in parallel, where probing is defined as the technique of perturbing measurements for identification, *i.e.*, finding the unknown locations of spoofed PMUs and narrowing down the searching ranges of the GSA phase shifts. The attacked PMU locations are determined at this stage, regardless of whether this PMU coordinates with other PMUs or not, *i.e.*, the proposed algorithm is robust against coordinated attacks.

**Definition 1** (Test dataset for identification on PMU $i$). *Define $N_i = M_i \cup S$ as the test dataset for the identification on PMU $i$, where $M_i$ is the set of the measurements from PMU $i$, $i \in \{2, ..., N_{pmu}\}$, and $S$ denotes the secure measurement set from the substation PMU and smart meters.*

There are only two remaining PMUs in $N_i$ besides all smart meters. Also, due to the existence of the secure substation PMU, there is at most one PMU under GSAs in $N_i$, and $N_i$ as the subset of all measurements still meets the system observability due to the radial nature of distribution systems.

Based on (20), the measurement residual between system variables and the measurements in $N_i$ is expressed as:

$$r_0 = K^\rho z^\rho = K^\rho [z_V^\rho \ z_I^\rho \ z_S]^T \quad (21)$$

where $z_V^\rho$ and $z_I^\rho$ are the voltage and current vectors measured by the substation PMU and PMU $i$, and $z_V^\rho = \begin{bmatrix} z_{V,slack} \\ z_{V,k} \end{bmatrix}$, $z_I^\rho = \begin{bmatrix} z_{I,slack} \\ z_{I,p} \end{bmatrix}$, and the vector elements in $z_V^\rho$ and $z_I^\rho$ are defined in Section II-B; $K^\rho$ denotes the residual sensitivity matrix of $N_i$.

GSAs introduce $\theta_i^{spf}$ to the data measured by PMU $i$, and the measurement residual under the GSA is expressed as

$$r_{spf} = K^\rho [z_V' \ z_I' \ z_S]^T \quad (22)$$

where $z_V'$ and $z_I'$ denote the voltage and current measurement

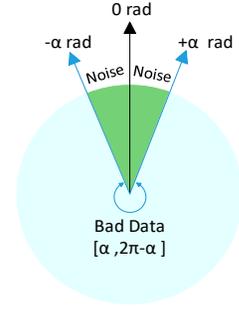

Fig. 2. Schematic diagram of phase measurement noise and bad data

TABLE I
TRUTH TABLE OF IDENTIFICATION METHOD

| PMU $i$ | $\theta_i^{spf}$ | $J_i^+$ | $J_i^-$ | $a_i$ |
|---|---|---|---|---|
| No GSAs | $\theta_i^{spf} = 0$ | ↑ | ↑ | 0 |
| Under GSAs | $\theta_i^{spf} \in (0, \pi)$ | ↑ | ↓ | 1 |
| | $\theta_i^{spf} = \pi$ | ↓ | ↓ | |
| | $\theta_i^{spf} \in (-\pi, 0)$ | ↓ | ↑ | |

↑ and ↓ indicate rising and descending residuals relative to $J_i$

vectors under this GSA, and $z_V' = \begin{bmatrix} z_{V,slack} \\ A_k z_{V,k} \end{bmatrix}$, $z_I' = \begin{bmatrix} z_{I,slack} \\ A_p z_{I,p} \end{bmatrix}$.

Substitute (22) into (2), and the sum of WMRs under this GSA is calculated by

$$J\left(\theta_i^{spf}\right) = r_{spf}^T w r_{spf} \quad (23)$$

where $w$ denotes the weight matrix of $N_i$, and $J\left(\theta_i^{spf}\right)$ depends on the only variable $\theta_i^{spf}$.

**Theorem 1** (Unimodality of $J\left(\theta_i^{spf}\right)$). *For $\theta_i^{spf} \in [-\pi, \pi]$, $J\left(\theta_i^{spf}\right)$ is a unimodal function with respect to $\theta_i^{spf}$. With $\theta_i^{spf} = 0$, i.e., there is no attacks, $J\left(\theta_i^{spf}\right)$ has a unique minimum $J_0$, and $J_0 = r_0^T w r_0$.*

*Proof*: See the appendix.

By designing the test datasets $N_i$, the potential coordinated attacks in the original measurement set transform into uncoordinated single GSA attacks on $N_i$, which are easier to detect. Also, **Theorem 1** is built on $N_i$ and conveniently used for the subsequent probing technique.

The proposed probing technique is described that a pair of phase angles $\pm\Delta\theta(\Delta\theta > 0)$ as probing signals are superposed onto $N_i$ to generate auxiliary measurement sets for the subsequent identification. The auxiliary measurement sets $N_i^+$ and $N_i^-$ are expressed as $\{z_V^+, z_I^+, z_S\}$ and $\{z_V^-, z_I^-, z_S\}$, where $z_V^+ = A^+ z_V'$, and $z_I^+ = A^+ z_I'$; $z_V^- = A^- z_V'$, and $z_I^- = A^- z_I'$; $A^+ = \begin{bmatrix} \cos\Delta\theta & -\sin\Delta\theta \\ \sin\Delta\theta & \cos\Delta\theta \end{bmatrix}$, and $A^- =$

$\begin{bmatrix} \cos(-\Delta\theta) & \sin\Delta\theta \\ \sin(-\Delta\theta) & \cos(-\Delta\theta) \end{bmatrix}$; The phase index $\varphi$ is suppressed for simplicity.

Suppose $J_i = J(\theta_i^{spf})$, $J_i^+ = J(\theta_i^{spf} + \Delta\theta)$, and $J_i^- = J(\theta_i^{spf} - \Delta\theta)$, and $J_i^+$ and $J_i^-$ are calculated by (2) based on $\boldsymbol{N}_i^+$ and $\boldsymbol{N}_i^-$, respectively. Further, two identification corollaries are used to obtain the truth table shown in Table I, where $a_i$ is a binary indicator, and $a_i = 1$ if PMU $i$ is attacked, and $a_i = 0$ otherwise. Specifically, **Corollary 1** is used to identify the locations of non-attack PMUs, while **Corollary 2** narrows down the ranges of phase shifts at attacked PMUs.

**Corollary 1.** *For PMU $i$, if $\theta_i^{spf} = 0$ and $0 < 2\alpha < \Delta\theta < 2\pi - 2\alpha$, then $J_i^+ > J_i$ and $J_i^- > J_i$. Here, $\alpha$ is the known maximum phase measurement error required by the IEEE Standard for PMU accuracy, i.e., 0.01 rad* [31].

*Proof*: Based on **Theorem 1**, if $0 < 2\alpha < \Delta\theta < 2\pi - 2\alpha$, the following characteristics hold:

$$\begin{cases} \alpha < \theta_i^{spf} + \Delta\theta + e_i < 2\pi - \alpha \\ \alpha - 2\pi < \theta_i^{spf} - \Delta\theta + e_i < -\alpha \end{cases} \leftrightarrow J_i^+ > J_i, J_i^- > J_i$$

where $\theta_i^{spf} = 0$, and $e_i$ denotes the inherent phase measurement noise on PMU $i$ and obeys a known Gaussian distribution, $-\alpha < e_i < \alpha$. The schematic diagram of noises and bad data is shown in Fig. 2. ∎

**Remark 1.** *In the case that PMU $i$ is not attacked, the probing is equivalent to launching an extra "GSA attack" on this non-attack PMU, rather than a noise-level signal. Such a probing signal leads to a larger measurement residual, and therefore, $\Delta\theta > 2\alpha$ is required.*

**Corollary 2.** *For PMU $i$, if $\theta_i^{spf} \in (0, \pi)$, then $J_i^+ > J_i$ and $J_i^- < J_i$; if $\theta_i^{spf} \in (-\pi, 0)$, then $J_i^+ < J_i$ and $J_i^- > J_i$; if $\theta_i^{spf} = \pi$, then $J_i^+ < J_i$ and $J_i^- < J_i$.*

*Proof*: Combined with $J(\theta_i^{spf}) \geq J_i$ in $[-\pi, \pi]$ and $J_i = J_0$ at $\theta_i^{spf} = 0$, $J(\theta_i^{spf})$ monotonically increases in $[0, \pi]$, while monotonically decreasing in $[-\pi, 0]$.

According to the monotonicity of $J(\theta_i^{spf})$ in $\theta_i^{spf} \in (0, \pi)$, we have

$$\pi > \theta_i^{spf} + \Delta\theta > \theta_i^{spf} > \theta_i^{spf} - \Delta\theta > 0 \leftrightarrow J_i^+ > J_i > J_i^-$$

The opposite occurs for $\theta_i^{spf} \in (-\pi, 0)$.

If $\theta_i^{spf} = \pi$, $J_i = J(\pi) = J(-\pi)$, and it is derived that

$$0 < \pi - \Delta\theta < \pi \leftrightarrow J(\pi - \Delta\theta) < J(\pi) \leftrightarrow J_i^- < J_i$$

$$-\pi < \Delta\theta - \pi < 0 \leftrightarrow (\Delta\theta - \pi) = J(\pi + \Delta\theta) < J(\pi)$$
$$\leftrightarrow J_i^+ < J_i \quad ∎$$

---

**Identification Procedure**

1 **Inputs**: Measurements $\boldsymbol{S}$ and $\boldsymbol{M}_i$.
2 **Parameters**: System model, $\boldsymbol{A}^+$, and $\boldsymbol{A}^-$.
3 **Initialization**: $\boldsymbol{P}_1 = \{\}, \boldsymbol{P}_2 = \{\}, \boldsymbol{P}_3 = \{\}$
4 **for** each PMU location $i$ **do**
5     $\boldsymbol{N}_i \leftarrow \boldsymbol{S} \cup \boldsymbol{M}_i$
6     Calculate $J_i$, $J_i^+$, and $J_i^-$.
7     **if** $J_i^+/J_i > 1$ && $J_i^-/J_i > 1$
      $\theta_i^{spf} = 0$
    **else if** $J_i^+/J_i < 1$ && $J_i^-/J_i < 1$
      $\boldsymbol{P}_1 \leftarrow \boldsymbol{P}_1 \cup \{i\}$ and $\theta_i^{spf} = \pi$
    **else if** $J_i^+/J_i > 1$ && $J_i^-/J_i < 1$
      $\boldsymbol{P}_2 \leftarrow \boldsymbol{P}_2 \cup \{i\}$ and $\theta_i^{spf} \in (0, \pi)$
    **else if** $J_i^+/J_i < 1$ && $J_i^-/J_i > 1$
      $\boldsymbol{P}_3 \leftarrow \boldsymbol{P}_3 \cup \{i\}$ and $\theta_i^{spf} \in (-\pi, 0)$
    **end**
8 **end**
9 **return** $\boldsymbol{P}_1$, $\boldsymbol{P}_2$, and $\boldsymbol{P}_3$

---

**Remark 2.** *The size of the probing signal $\Delta\theta$ in Corollary 2 requires meeting $\Delta\theta < \theta_i^{spf}$, which is not difficult to achieve, since these phase shifts are usually significant* [11], [13]. *Typical spoofed phase angles such as 52°, 60° and 70° are reported in* [5], [8], *and* [9]. *For instance, in* [5], *a GSA model is formulated by maximizing the difference $\Delta t_i$ between the spoofed clock offset and the pre-attack clock offset, which results in the maximum GSA phase angle according to $\theta_i^{spf} = 2\pi f \Delta t_i$. Also, if $\theta_i^{spf} < \Delta\theta$, the original influence of $\theta_i^{spf}$ on $J(\theta_i^{spf})$ may be neutralized, even overtaken by $\Delta\theta$. In such a case, whether the identification results are caused by $\theta_i^{spf}$ or $\Delta\theta$ cannot be determined. As a result, the identification method fails to reflect the true interval that $\theta_i^{spf}$ falls in.*

In order to successfully realize the above identification, it is concluded that $0 < 2\alpha < \Delta\theta \ll |\theta_i^{spf}|$ should hold. The identification procedure is shown in the following pseudo-code, where $\boldsymbol{P}_1$, $\boldsymbol{P}_2$, and $\boldsymbol{P}_3$ represent the category sets of these identified PMUs, and $\boldsymbol{P}_1 = \{i|\theta_i^{spf} = \pi\}$, $\boldsymbol{P}_2 = \{i|\theta_i^{spf} \in (0, \pi)\}$, $\boldsymbol{P}_3 = \{i|\theta_i^{spf} \in (-\pi, 0)\}$.

*D. Correction of Spoofed PMU Data*

The spoofed PMU locations are determined by the above identification method, and the searching scale of each GSA phase shift is explicitly determined according to Table I. With the knowledge of the locations and ranges of $\theta_i^{spf}$, the correction algorithm introduces the inverse matrix of $\boldsymbol{A}$ to minimize the mismatch between system states and all measurements.

The inverse matrix of $\boldsymbol{A}$ could be obtained with all non-zero $\theta_i^{spf}$ estimated, and the measurement vector is corrected as:

$$\boldsymbol{z}_{corr} = \boldsymbol{A}^{-1}\boldsymbol{z}_{spf} \quad (24)$$

where $\boldsymbol{A}^{-1}$ is also a block diagonal matrix and composed of block submatrices according to the property of the block diagonal matrix $\boldsymbol{A}$. The blocks associated with PMU $i$ in $\boldsymbol{A}^{-1}$ are expressed as

$$(\boldsymbol{A}_i^\varphi)^{-1} = \begin{bmatrix} \cos\theta_i^{spf} & \sin\theta_i^{spf} \\ -\sin\theta_i^{spf} & \cos\theta_i^{spf} \end{bmatrix}$$

The measurement residual with the correction of all PMU measurements is obtained by

$$\boldsymbol{r}_{corr} = \boldsymbol{z}_{corr} - \boldsymbol{H}\boldsymbol{x}_{corr} \quad (25)$$

where $\boldsymbol{x}_{corr}$ as the corrected state vector is accurately obtained by the iterative estimation process in Section II-A.

The sum of WMRs is calculated as:

$$J_{corr} = \boldsymbol{r}_{corr}^T \boldsymbol{W} \boldsymbol{r}_{corr} = f\left(\theta_1^{spf}, \theta_2^{spf}, \ldots, \theta_{N_{pmu}}^{spf}\right) \quad (26)$$

These phase shifts are obtained by minimizing $J_{corr}$ to recover the system states:

$$\left(\theta_1^{spf}, \theta_2^{spf}, \ldots, \theta_{N_{pmu}}^{spf}\right) = \arg\min J_{corr} \quad (27.a)$$

$$s.t. \quad \theta_i^{spf} = 0 \quad \forall i \notin \boldsymbol{P}_1 \cup \boldsymbol{P}_2 \cup \boldsymbol{P}_3 \quad (27.b)$$

$$\theta_i^{spf} = \pi \quad \forall i \in \boldsymbol{P}_1 \quad (27.c)$$

$$\theta_i^{spf} = (0, \pi) \quad \forall i \in \boldsymbol{P}_2 \quad (27.d)$$

$$\theta_i^{spf} = (-\pi, 0) \quad \forall i \in \boldsymbol{P}_3 \quad (27.e)$$

where $\boldsymbol{P}_1$, $\boldsymbol{P}_2$, and $\boldsymbol{P}_3$ are determined in advance.

The solution of (27) enables the recovery of the phase angles from spoofed PMUs. Most optimization problems in DSSE are non-convex, considering the quadratic characteristics of objective functions, which are established by a WLS criterion, and the nonlinearity of the measurement residual about decision variables in these optimization models. Consequently, heuristic search algorithms are widely used in DSSE, such as [32] – [34]. Particle swarm optimization (PSO) is employed to solve (27), and the variables in (27.d) and (27.e) as particles float in the corresponding ranges. The detailed implementation of PSO can be found in [34, Chapter 4].

It is noted that according to the identification results in (27.b-e), the amount of unknown variables in (27) decreases. The GSA phase shifts on PMU $i$ ($i \in \boldsymbol{P}_2 \cup \boldsymbol{P}_3$) are reserved as unknown variables, while the values of the remaining phase shifts are fixed as 0 or $\pi$. The risk of the solution being trapped in a local minima is reduced by: 1) the proposed identification method, which narrows down the searching locations and ranges of GSA phase shifts prior to the subsequent correction algorithm, 2) increasing the amount of particles in the PSO algorithm and the maximum iteration times [34].

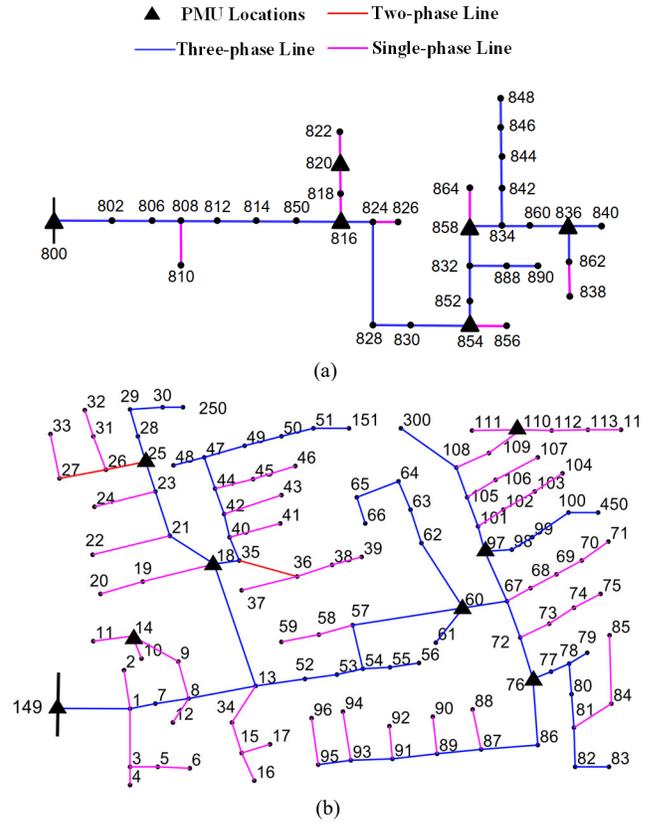

Fig. 3. Schematic diagrams of (a) the 34-bus system (b) the 123-bus system.

TABLE II
PMU PLACEMENT LOCATIONS

| System Scale | Buses with PMUs | Measurement Redundancy |
|---|---|---|
| 34-bus | 800, 816, 820, 836, 854, 858 | 1.324 |
| 123-bus | 149, 14, 18, 25, 60, 76, 97, 110 | 1.126 |

IV. SIMULATION TEST

The proposed algorithm is tested on the unbalanced IEEE 34-bus, 24.9 kV and 123-bus, 4.16 kV distribution networks [35], shown in Fig.3. The nominal capacities of the 34-bus and 123-bus systems are 2500 kVA and 5000 kVA, respectively. The following conditions are applied to the maximum errors of measurements that obey Gaussian distributions, and the standard deviation of a particular measurement is one-third of the corresponding maximum error [24]:

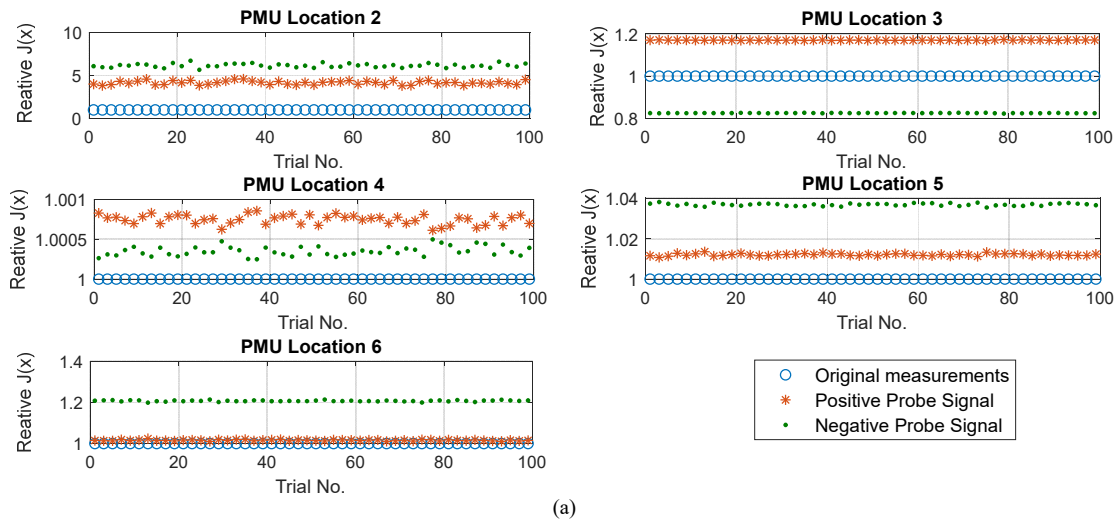

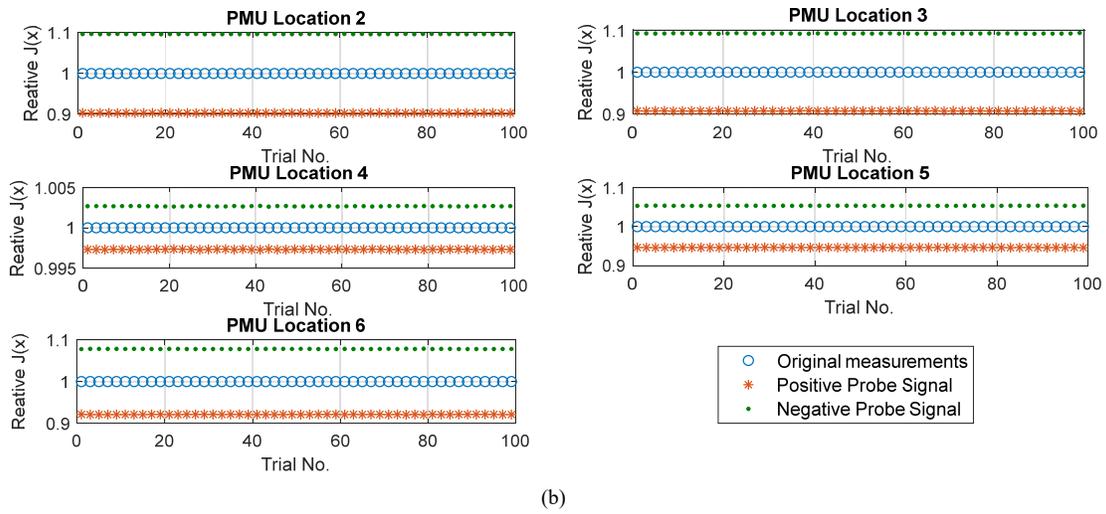

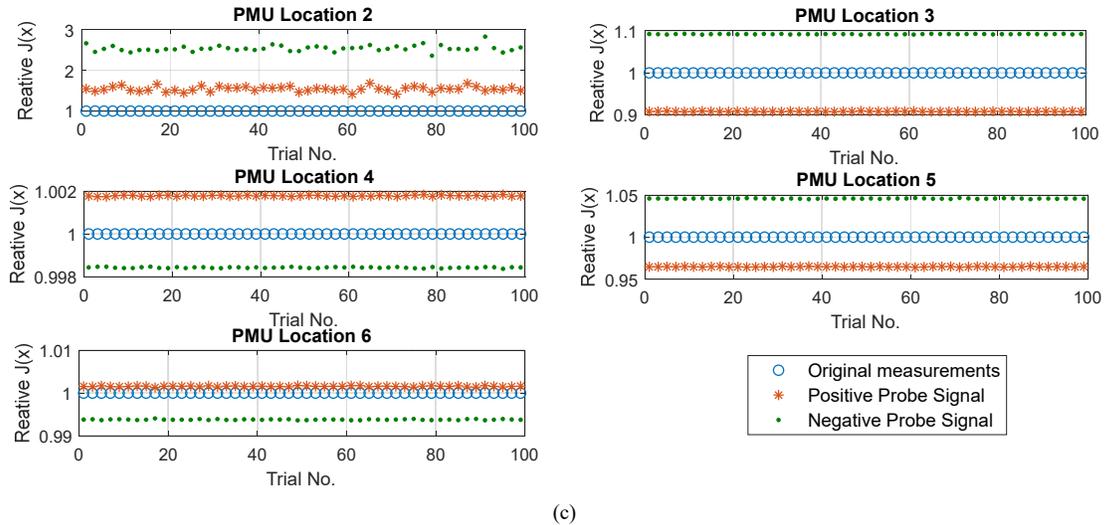

Fig. 4. Identification results (a) in Scenario 1. (b) in Scenario 2. (c) in Scenario 3.

1) PMU measurements: 1% of true values of voltages and currents for magnitudes and 1 crad ($10^{-2}$ rad) for phase angles [31]. Table II and Fig.3 give the PMU placement profile.

2) Smart meter measurements: 3% of true values for active and reactive powers. The smart meters, which become available for distribution grids owing to their cheap costs, are assumed to be installed at each load bus [36].

All test cases run for 100 times of Monte Carlo trials. Three types of scenarios are tested.

Scenario 1: Single GSA

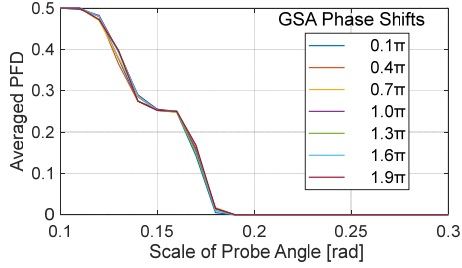

Fig. 5. PFD of the identification method

TABLE III
ESTIMATION ERRORS IN THE 34-BUS SYSTEM

| Estimation Errors [rad] | | Scenario | | |
| --- | --- | --- | --- | --- |
| | | 1 | 2 | 3 |
| GSA Phase Shift [rad] | 0.1π | 0.0110 | 0.1127 | 0.0505 |
| | 0.4π | 0.0090 | 0.0211 | 0.0216 |
| | 0.7π | 0.0122 | 0.0218 | 0.0209 |
| | 1.0π | 0 | 0 | 0.0154 |
| | 1.3π | 0.0095 | 0.0266 | 0.0210 |
| | 1.6 π | 0.0116 | 0.0223 | 0.0217 |
| | 1.9 π | 0.0092 | 0.1112 | 0.0410 |

TABLE IV
ESTIMATION ERRORS IN THE 123-BUS SYSTEM

| Estimation Errors [rad] | | Scenario | | |
| --- | --- | --- | --- | --- |
| | | 1 | 2 | 3 |
| GSA Phase Shift [rad] | 0.1π | 0.0109 | 0.1120 | 0.1115 |
| | 0.4π | 0.0106 | 0.0137 | 0.0179 |
| | 0.7π | 0.0112 | 0.0245 | 0.0207 |
| | 1.0π | 0 | 0 | 0.0080 |
| | 1.3π | 0.0046 | 0.0241 | 0.0148 |
| | 1.6 π | 0.0048 | 0.0174 | 0.0191 |
| | 1.9 π | 0.0068 | 0.1319 | 0.0607 |

Scenario 2: Multiple GSAs with same phase shifts
Scenario 3: Multiple GSAs with different phase shifts

Scenarios 1 and 2 are jointly designed to illustrate that the proposed algorithm is applicable to not only a single GSA but also multiple GSAs. Scenarios 2 and 3 are jointly designed to verify that this algorithm is not influenced by the magnitudes of GSA phase shifts. Define $\psi$ as a set of GSA phase shifts in all scenarios, and the first element in $\psi$ is always zero, since the substation PMU is immune to GSAs.

### A. Identification of Multiple GSAs

To verify the correctness of the identification method, three types of scenarios in the 34-bus system are designed as follows:
Scenario 1: $\psi$ (0, 0, 0.5π, 0, 0, 0)
Scenario 2: $\psi$ (0, -0.5π, -0.5π, -0.5π, -0.5π, -0.5π)
Scenario 3: $\psi$ (0, 0, -0.5π, 0.2π, -0.1π, 4.9π)

In each scenario, $J_i^+/J_i$ and $J_i^-/J_i$ on PMU $i$, respectively corresponding to the positive and negative probing signals, are compared with 1, and the identification results in 100 Monte Carlo trials are shown in Fig. 4. In Fig. 4(a), the GSA location on the 3rd PMU is identified correctly, and the phase shift in this GSA is identified in (0, π). Moreover, all GSA locations in Scenarios 2 and 3 are correctly identified. In Scenario 3, although the last shift angle 4.9π is out of the interval [-π, π], its effect on DSSE in rectangular coordinates is equivalent to that of 0.9π. To sum up, the identification method is not influenced by the amount and the locations of multiple GSAs.

### B. Identification Sensitivity

Different combinations of probing signals and GSA phase shifts are tested to investigate the sensitivity of the identification method. Two indices, the probability of miss detection (PMD) and the probability of false detection (PFD), are utilized to evaluate the sensitivity of the identification method. PMD describes the probability that the algorithm fails in finding the locations of attacked PMUs, and PFD describes the probability that the non-attack PMUs are misjudged as maliciously spoofed. With the values of both indices closer to zero, this algorithm has a better identification performance.

The average PMD and PFD are calculated as follows:

$$\text{PMD} = \frac{1}{N_{tr}\sum_{i=2}^{N_{pmu}} a_i} \sum_{j=1}^{N_{tr}} \sum_{i=2}^{N_{pmu}} (a_i \alpha_{ij}) \quad (28.a)$$

$$\text{PFD} = \frac{1}{N_{tr}\sum_{i=2}^{N_{pmu}}(1-a_i)} \sum_{j=1}^{N_{tr}} \sum_{i=2}^{N_{pmu}} [(1-a_i)\beta_{ij}] \quad (28.b)$$

where $\alpha_{ij}$ and $\beta_{ij}$ represent the amount of miss detection and false detection at PMU $i$ in the $j$th trial, respectively; $N_{tr}$ is the total amount of trials, and $a_i$ is a binary indicator as before.

These simulations show that PMDs in different combinations of the scale of probing signals and GSA phase shifts are always equal to zero. Fig. 5 depicts the PFD trends, where different combinations of $\Delta\theta$ and GSA phase shifts are tested in Scenario 1. When $\Delta\theta$ is larger than 0.18 rads, the algorithm avoids misjudging the non-attack PMUs as attacked for all GSA phase shifts. The measurement errors of other meters (*i.e.*, smart meters) cause these non-zero PFDs in Fig.5 and these errors are inevitable and have coordinated impacts along with PMU phase errors on the DSSE accuracy. In conclusion, the proposed algorithm has excellent performance in avoiding miss and false identification when the size of $\Delta\theta$ is appropriately selected. The threshold of $\Delta\theta$ can be obtained by prior knowledge and/or off-line probing tests based on the proposed identification algorithm. With the development of micro-PMUs with higher measurement accuracy (*e.g.*, 0.003 degrees for the maximum error of phase angle [37]), the proposed method could provide a wider range for the sizes of $\Delta\theta$ to choose.

### C. Correction Accuracy

The average estimation errors of GSA phase shifts are used to evaluate the correction accuracy for multiple GSAs. They are calculated as

$$\varepsilon = \frac{1}{N_{tr}\sum_{i=2}^{N_{pmu}} a_i} \sum_{j=1}^{N_{tr}} \sum_{i=2}^{N_{pmu}} (a_i |\Delta\theta_{tr,ij} - \Delta\theta_{est,ij}|) \quad (29)$$

where $\Delta\theta_{tr,ij}$ and $\Delta\theta_{est,ij}$ are the true and estimated values of the phase shift on PMU $i$ in the $j$th trial.

In the 34-bus system, three types of scenarios are designed as
Scenario 1: $\psi\left(0, \theta_2^{spf}, 0, 0, 0, 0\right)$

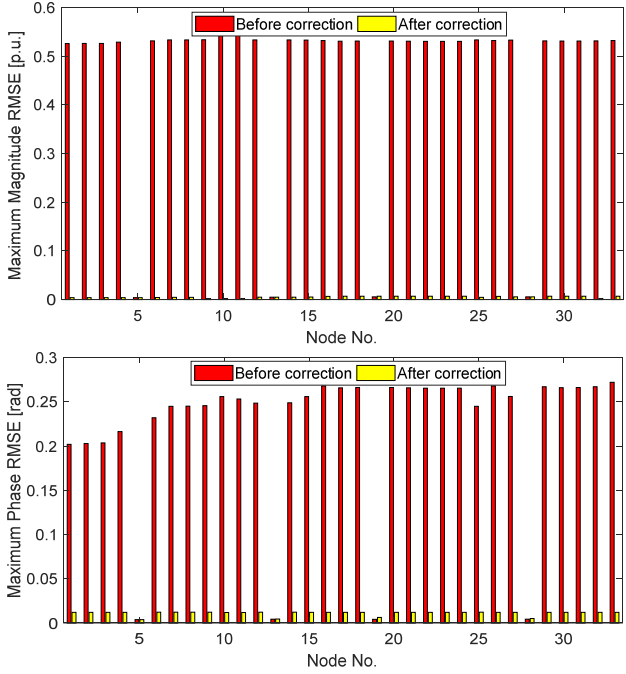
Fig. 6. Comparison of RMSEs before and after corrections

TABLE V
RMSEs ON PHASE $a$ OF DIFFERENT GSAs IN 34-BUS SYSTEM

| Average RMSEs | Scenario | | |
|---|---|---|---|
| | 1 | 2 | 3 |
| $E(\delta_{V,k}^A)$ [p.u.] | 0.0139 | 0.0158 | 0.0157 |
| $E(\delta_{\theta,k}^A)$ [rad] | 0.0087 | 0.0143 | 0.0181 |

Scenario 2: $\psi\left(0, \theta_2^{spf}, \theta_2^{spf}, \theta_2^{spf}, \theta_2^{spf}, \theta_2^{spf}\right)$
Scenario 3: $\psi\left(0, -0.5\pi, -0.5\pi, -0.5\pi, \theta_5^{spf}, -0.5\pi\right)$

The phase shifts $\theta_2^{spf}$ and $\theta_5^{spf}$ are set to vary from $0.1\pi$ to $1.9\pi$, and the step is $0.3\pi$. Table III summarizes the estimation errors of GSA phase shifts in three scenarios. It is clear from Table III that these multiple GSAs with different locations and magnitudes are accurately corrected. The estimation results in the 123-bus system are shown in Table IV, which verifies the robustness of this algorithm when the network scale increases.

### D. DSSE Performance

The goal of attack correction under GSAs is to provide reliable measurements for power system monitoring and control. Therefore, the root mean square errors (RMSEs) of variables are utilized to evaluate the estimation performance. The RMSEs of voltage magnitudes and phase angles at each node are calculated by

$$\delta_{V,k}^\varphi = \sqrt{E\left(\left(\frac{\tilde{V}_k^\varphi - V_k^\varphi}{\tilde{V}_k^p}\right)^2\right)} \tag{30.a}$$

$$\delta_{\theta,k}^\varphi = \sqrt{E\left(\left(\tilde{\theta}_k^\varphi - \theta_k^\varphi\right)^2\right)} \tag{30.b}$$

TABLE VI
COMUTATATION TIME IN DIFFERENT SCENARIOS

| Average Computation Time [s] | | Scenario | | |
|---|---|---|---|---|
| | | 1 | 2 | 3 |
| GSA Phase Shift [rad] | $0.1\pi$ | 1.928 | 79.48 | 74.60 |
| | $0.4\pi$ | 1.734 | 65.81 | 67.34 |
| | $0.7\pi$ | 1.690 | 72.47 | 69.11 |
| | $1.0\pi$ | 0.643 | 0.604 | 66.12 |
| | $1.3\pi$ | 1.609 | 75.02 | 70.59 |
| | $1.6\pi$ | 1.656 | 68.43 | 68.11 |
| | $1.9\pi$ | 1.644 | 75.66 | 75.36 |

TABLE VII
DG PLACEMENTS IN 123-BUS SYSTEM

| Type | No. node | Capacity [kW/ph] | Phase | Type | No. node | Capacity [kW/ph] | Phase |
|---|---|---|---|---|---|---|---|
| PV | 14 | 200 | $c$ | Wind | 250 | 300 | $a,b,c$ |
| Wind | 61 | 300 | $a,b,c$ | PV | 300 | 200 | $a$ |
| Wind | 151 | 300 | $a,b,c$ | PV | 450 | 200 | $a$ |

where $V_k^\varphi$ and $\theta_k^\varphi$ denote the voltage magnitude and phase angle on phase $\varphi$ at node $k$, and $\tilde{V}_k^\varphi$ and $\tilde{\theta}_k^\varphi$ are the corresponding true values; $E(\cdot)$ is the expected value of these variables.

Scenario 3 is tested on the 34-bus system, where the GSA phase shifts are set as $\psi(0, 0, -0.5\pi, 0.2\pi, -0.1\pi, 4.9\pi)$ to generate the spoofed measurements. Further, the maximum of RMSEs in three-phase voltages obtained by (30) are selected and shown in Fig.6 to evaluate the estimation accuracy of the proposed correction algorithm. Also, Table V lists the average A-phase RMSEs in these scenarios for various GSAs used in Section IV-C. It is shown that the proposed method enables the improvement of the DSSE accuracy in both magnitudes and phase angles by correcting multiple GSAs.

### E. Computation Time

Experiments for different scenarios are carried out to illustrate the computational efficiency of the proposed algorithm. They are implemented on a PC with 2.6 GHz i5, and 8GB RAM using Matlab 2017b.

The average computation time for various scenarios is listed in Table VI, where the scenario design is the same as the one in Section IV-C. These results indicate that the proposed algorithm rapidly realizes the identification and correction of multiple GSAs.

### F. Impacts of DG penetration

The impacts of DGs on the proposed algorithm are considered by adding DGs into the 34-bus and 123-bus distribution systems. The power outputs of DGs are measured by smart meters, and all DGs are modeled as PQ buses with a constant power factor of 0.95 [33]. Two wind units are installed on the phase $a$ of nodes 802 and 822 in the 34-bus system, and the installed capacity of each DG is 200 kW. Moreover, six DGs including PV and wind units are considered in the 123-bus system, and Table VII lists the installation details. Also, the total power outputs of DGs account for about 80% of total load demands.

TABLE VIII
COMPUTATIONAL EFFICIENCY IN ACTIVE DISTRIBUTION SYSTEMS

| GSA Phase Shifts [rad] | 34-bus System with DGs | | 123-bus System with DGs | |
|---|---|---|---|---|
| | Time [s] | Error [rad] | Time [s] | Error [rad] |
| $\theta_3^{spf} = 0.05\pi$ | 1.58 | 0.0308 | 333.74 | 0.0316 |
| $\theta_2^{spf} = 0.1\pi$ | 1.64 | 0.0110 | 337.18 | 0.0117 |
| $\theta_2^{spf} = \frac{\pi}{3}, \theta_5^{spf} = 0.5\pi$ | 45.03 | 0.0148 | 521.34 | 0.0151 |

TABLE IX
COMPARISON IN THE 34-BUS SYSTEM

| $\psi(0,\theta_2^{spf},0,0,0,0)$ | | Estimation Errors [rad] | | Computing Time [s] | |
|---|---|---|---|---|---|
| | | Proposed Method | [11] | Proposed Method | [11] |
| GSA Phase Shift [rad] | $0.1\pi$ | 0.0110 | 0.0280 | 1.928 | 12.932 |
| | $0.4\pi$ | 0.0090 | 0.0288 | 1.734 | 13.394 |
| | $0.7\pi$ | 0.0122 | 0.0288 | 1.690 | 13.972 |
| | $1.0\pi$ | 0 | 0.0271 | 0.643 | 12.384 |
| | $1.3\pi$ | 0.0095 | 0.0273 | 1.609 | 12.986 |
| | $1.6\pi$ | 0.0116 | 0.0270 | 1.656 | 13.394 |
| | $1.9\pi$ | 0.0092 | 0.0251 | 1.644 | 12.846 |

Table VIII shows the computation time and estimation errors of GSA phase shifts, and a smaller-size GSA phase shift, $0.05\pi$, is tested. These results show that the high accuracy of the proposed algorithm still firmly holds with the DG penetration.

### G. Comparison with Other Methods

The proposed method enables the correction of multiple GSAs in unbalanced distribution systems, while [11]–[13] are applied to transmission systems. As mentioned in Introduction, the existing studies on the cyber-attacks in transmission systems cannot be trivially extended to distribution systems. Also, [11]–[13] are built on the generalized likelihood ratio tests, where all combination of the attacked PMU locations are attempted to detect the presence of a GSA. For instance, in [11], based on the assumption of a single GSA, the best $\theta_i^{spf}$ for each PMU is estimated in $[0, 2\pi)$, and the phase shifts of other PMUs except PMU $i$ are set as zero. The global estimation $\hat{\theta}_i^{spf}$ in [11] is obtained by

$$\hat{\theta}_i^{spf} = \arg \min_{\theta_i^{spf} \in [0,2\pi)} J_{corr} \qquad i = 1, \ldots, N_{pmu} \quad (31)$$

As [11] one-dimensionally searches $\theta_i^{spf}$ through all PMUs, the proposed probing method restricts the location of $\theta_i^{spf}$ on the exact attacked PMU efficiently. Scenario 1 is tested on the 34-bus distribution system to demonstrate the computational efficiency of the proposed algorithm. The estimation results of this algorithm are compared with the golden section search from [11] in Table. IX. In the golden section search, there is no identification procedure for the attacked PMUs prior to this search in [11]. It is noted that the proposed algorithm does not rely on the assumption of a single GSA owing to the proposed probing algorithm, and this GSA is determined as a single GSA at the identification stage. Moreover, the location of this attacked PMU is determined, and the ranges of this phase shift on the attacked PMU shrink. Consequently, both the efficiency and accuracy of the proposed algorithm are improved.

## V. CONCLUSION

This paper proposes a novel algorithm for identifying and correcting PMU data under multiple GSAs in distribution systems. In contrast to a brute force search for all combinations of spoofed PMU locations and magnitudes of GSA phase shifts, which is prohibitive, the proposed algorithm is hierarchical. Moreover, the proposed identification method not only determines the locations of multiple spoofed PMUs but also successfully narrows down the scale for these phase shifts. Consequently, the computational efficiency of the subsequent correction for GSAs is improved. Future work focuses on the extension of this algorithm to false data injection attacks.

## APPENDIX

*Proof of Theorem 1:* The closed-form solution of the estimated states used in [21] is used to prove the unimodality of $J\left(\theta_i^{spf}\right)$ based on $N_i$ for the identification at PMU $i$. Given $J_0 = (K^\rho z^\rho)^T w(K^\rho z^\rho)$, where $w$ is a diagonal matrix, let $M = (K^\rho)^T w K^\rho$, and $M$ is a symmetric matrix. Reorganize $M$ in a block form, and $J_0$ and $J\left(\theta_i^{spf}\right)$ are calculated by

$$J_0 = \begin{bmatrix} z_V^\rho \\ z_I^\rho \\ z_S \end{bmatrix}^T \begin{bmatrix} M_{11} & M_{12} & M_{13} \\ M_{21} & M_{22} & M_{23} \\ M_{31} & M_{32} & M_{33} \end{bmatrix} \begin{bmatrix} z_V^\rho \\ z_I^\rho \\ z_S \end{bmatrix} \quad (32)$$

$$J\left(\theta_i^{spf}\right) = \begin{bmatrix} z_V' \\ z_I' \\ z_S \end{bmatrix}^T \begin{bmatrix} M_{11} & M_{12} & M_{13} \\ M_{21} & M_{22} & M_{23} \\ M_{31} & M_{32} & M_{33} \end{bmatrix} \begin{bmatrix} z_V' \\ z_I' \\ z_S \end{bmatrix} \quad (33)$$

Define $\Delta J\left(\chi, \theta_i^{spf}\right) = J\left(\theta_i^{spf}\right) - J_0$, where $\chi$ denotes the set of the phase angles influenced by $\theta_i^{spf}$, $\chi = \{\theta_{V,k}^a, \theta_{V,k}^b, \theta_{V,k}^c, \theta_{I,p}^a, \theta_{I,p}^b, \theta_{I,p}^c\}$. It is deduced that

$$\Delta J\left(\chi, \theta_i^{spf}\right) =$$
$$\begin{bmatrix} z_V' \\ z_I' \end{bmatrix}^T \begin{bmatrix} M_{11} & M_{12} \\ M_{21} & M_{22} \end{bmatrix} \begin{bmatrix} z_V' \\ z_I' \end{bmatrix} - \begin{bmatrix} z_V^\rho \\ z_I^\rho \end{bmatrix}^T \begin{bmatrix} M_{11} & M_{12} \\ M_{21} & M_{22} \end{bmatrix} \begin{bmatrix} z_V^\rho \\ z_I^\rho \end{bmatrix} +$$
$$2 z_S^T B (\begin{bmatrix} z_V' \\ z_I' \end{bmatrix} - \begin{bmatrix} z_V^\rho \\ z_I^\rho \end{bmatrix}) = \sum_{j=1}^{4} F_j \quad (34)$$

$$F_1 = {z_V'}^T M_{11} z_V' - {z_V^\rho}^T M_{11} z_V^\rho \quad (35.a)$$
$$F_2 = 2{z_V'}^T M_{12} z_I' - 2{z_V^\rho}^T M_{12} z_I^\rho \quad (35.b)$$
$$F_3 = {z_I'}^T M_{22} z_I' - {z_I^\rho}^T M_{22} z_I^\rho \quad (35.c)$$
$$F_4 = 2 z_S^T B (\begin{bmatrix} z_V' \\ z_I' \end{bmatrix} - \begin{bmatrix} z_V^\rho \\ z_I^\rho \end{bmatrix}) \quad (35.d)$$

where $\boldsymbol{B} = [\boldsymbol{M}_{31}\ \boldsymbol{M}_{32}]$, and $\boldsymbol{M}_{12} = \boldsymbol{M}_{21}$. $\boldsymbol{z}'_V = \begin{bmatrix} |V^\varphi_{slack}|\cos\theta^\varphi_{V,slack} \\ |V^\varphi_{slack}|\sin\theta^\varphi_{V,slack} \\ |V^\varphi_k|\cos(\theta^\varphi_{V,k}+\theta^{spf}_i) \\ |V^\varphi_k|\sin(\theta^\varphi_{V,k}+\theta^{spf}_i) \end{bmatrix}$, $\boldsymbol{z}'_I = \begin{bmatrix} |I^\varphi_{slack}|\cos\theta^\varphi_{I,slack} \\ |I^\varphi_{slack}|\sin\theta^\varphi_{I,slack} \\ |I^\varphi_p|\cos(\theta^\varphi_{I,p}+\theta^{spf}_i) \\ |I^\varphi_p|\sin(\theta^\varphi_{I,p}+\theta^{spf}_i) \end{bmatrix}$; $\boldsymbol{z}^\rho_V = \begin{bmatrix} |V^\varphi_{slack}|\cos\theta^\varphi_{V,slack} \\ |V^\varphi_{slack}|\sin\theta^\varphi_{V,slack} \\ |V^\varphi_k|\cos\theta^\varphi_{V,k} \\ |V^\varphi_k|\sin\theta^\varphi_{V,k} \end{bmatrix}$, $\boldsymbol{z}^\rho_I = \begin{bmatrix} |I^\varphi_{slack}|\cos\theta^\varphi_{I,slack} \\ |I^\varphi_{slack}|\sin\theta^\varphi_{I,slack} \\ |I^\varphi_p|\cos\theta^\varphi_{I,p} \\ |I^\varphi_p|\sin\theta^\varphi_{I,p} \end{bmatrix}$.

Take $F_1$ as an example for expansion and further analysis, and without loss of generality, let $\boldsymbol{M}_{11} = \begin{bmatrix} \boldsymbol{m}_{11} & \boldsymbol{m}_{12} \\ \boldsymbol{m}_{21} & \boldsymbol{m}_{22} \end{bmatrix}$, and $\boldsymbol{m}_{12} = \boldsymbol{m}_{21}$, $\boldsymbol{m}_{22} = \begin{bmatrix} b_{11} & b_{12} \\ b_{21} & b_{22} \end{bmatrix}$. For simplicity, the auxiliary functions $f_1(\theta_V, \theta^{spf}_i)$ with the node index $k$ and phase index $\varphi$ suppressed is used in the following proof process:

$$f_1 = 2\boldsymbol{s}^T \boldsymbol{m}_{12}(\boldsymbol{y}' - \boldsymbol{y}) + \boldsymbol{y}'^T \boldsymbol{m}_{22}\boldsymbol{y}' - \boldsymbol{y}^T \boldsymbol{m}_{22}\boldsymbol{y} \quad (36)$$

where $\boldsymbol{s} = \begin{bmatrix} \cos\theta_{V,slack} \\ \sin\theta_{V,slack} \end{bmatrix}$, $\boldsymbol{y} = \begin{bmatrix} \cos\theta_V \\ \sin\theta_V \end{bmatrix}$, and $\boldsymbol{y}' = \begin{bmatrix} \cos(\theta_V + \theta^{spf}_i) \\ \sin(\theta_V + \theta^{spf}_i) \end{bmatrix}$.

Then, (36) is further expressed as

$$f_1 = f_{11} + f_{12} \quad (37)$$

where $f_{11} = 2\boldsymbol{s}^T \boldsymbol{m}_{12}(\boldsymbol{y}' - \boldsymbol{y}) = \lambda_1\big(\cos(\theta_V + \theta^{spf}_i) - \cos\theta_V\big) + \lambda_2\big(\sin(\theta_V + \theta^{spf}_i) - \sin\theta_V\big) + \lambda_3$, $\lambda_1$, $\lambda_2$, and $\lambda_3$ are constants about $\boldsymbol{s}$ and $\boldsymbol{m}_{12}$; $f_{12} = \boldsymbol{y}'^T \boldsymbol{m}_{22}\boldsymbol{y}' - \boldsymbol{y}^T \boldsymbol{m}_{22}\boldsymbol{y} = b_{11}\big[\cos^2(\theta_V + \theta^{spf}_i) - \cos^2\theta_V\big] + b_{22}\big[\sin^2(\theta_V + \theta^{spf}_i) - \sin^2\theta_V\big] + b_{12}\big[\sin(2\theta_V + 2\theta^{spf}_i) - \sin(2\theta_V)\big]$.

To find the extreme values of $f_1$, take a partial derivative of $f_1$ with respect to $\theta_V$, and use sum-to-product formulas in the following partial derivatives as:

$$\frac{\partial f_{11}}{\partial \theta_V} = \left[\lambda_1 \sin\left(\theta_V + \frac{\theta^{spf}_i}{2}\right) + \lambda_2 \cos\left(\theta_V + \frac{\theta^{spf}_i}{2}\right)\right] \sin\theta^{spf}_i \big/ \cos\frac{\theta^{spf}_i}{2}$$

$$\frac{\partial f_{12}}{\partial \theta_V} = [2(b_{22} - b_{11})\cos(2\theta_V + \theta^{spf}_i) + 4b_{12}\sin(2\theta_V + \theta^{spf}_i)]\sin\theta^{spf}_i$$

Hence, $\frac{\partial f_1}{\partial \theta_V}$ has the common factor $\sin\theta^{spf}_i$ as well as $\frac{\partial F_1}{\partial \theta_V}$ with the phase indices included. For the similar reason, $\sin\theta^{spf}_i$ is also the common factor in $\frac{\partial F_2}{\partial \theta_V}, \frac{\partial F_2}{\partial \theta_I}, \frac{\partial F_3}{\partial \theta_I}, \frac{\partial F_4}{\partial \theta_V}$, and $\frac{\partial F_4}{\partial \theta_I}$. Also, let $\frac{\partial \Delta J}{\partial \theta_V} = 0$ or $\frac{\partial \Delta J}{\partial \theta_I} = 0$, and solve $\theta^{spf}_i = 0$ within [-π, π], $\Delta J$ finds the unique minimum, i.e., 0, under any values of $\boldsymbol{\chi}$. Then, $J(\theta^{spf}_i) = J_0$ holds at $\theta^{spf}_i = 0$.

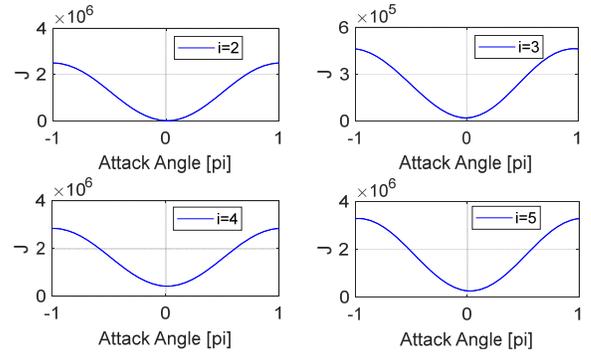

Fig. 7. $J(\theta^{spf}_i)$ in the 34-bus system shown in the case study, $i \in \{2, 3, 4, 5\}$, when the GSA $\psi(0, \theta^{spf}_2, \theta^{spf}_3, \theta^{spf}_4, \theta^{spf}_5, 0)$ occurs at multiple PMUs.

The calculation results of $J(\theta^{spf}_i)$ in (33) from an exhaustive search method are shown in Fig.7. It is also shown that $J(\theta^{spf}_i)$ is a unimodal function of $\theta^{spf}_i$.

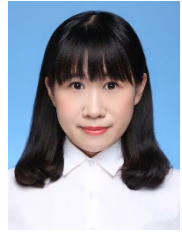

**Ying Zhang** (S'18) received the B.S. and M.S. degrees in electrical engineering from Shandong University, Jinan, China, in 2014 and 2017. She is currently pursuing the Ph.D. degree in Department of Electrical and Computer Engineering at Southern Methodist University, Dallas, Texas, USA. Her research interests include distribution system state estimation and its applications with PMU data.

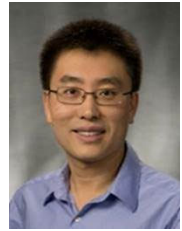

**Jianhui Wang** (M'07-SM'12) received the Ph.D. degree in electrical engineering from Illinois Institute of Technology, Chicago, Illinois, USA, in 2007. Presently, he is an Associate Professor with the Department of Electrical and Computer Engineering at Southern Methodist University, Dallas, Texas, USA. Prior to joining SMU, Dr. Wang had an eleven-year stint at Argonne National Laboratory with the last appointment as Section Lead – Advanced Grid Modeling. Dr. Wang is the secretary of the IEEE Power & Energy Society (PES) Power System Operations, Planning & Economics Committee. He has held visiting positions in Europe, Australia and Hong Kong including a VELUX Visiting Professorship at the Technical University of Denmark (DTU). Dr. Wang is the Editor-in-Chief of the IEEE Transactions on Smart Grid and an IEEE PES Distinguished Lecturer. He is also a Clarivate Analytics highly cited researcher for 2018.

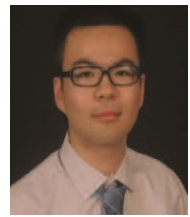

**Jianzhe Liu** (S'13-M'17) received the B.E. degree in electrical engineering from Huazhong University of Science and Technology, Wuhan, China, in 2012, and the Ph.D. degree in electrical and computer engineering from The Ohio State University, Ohio, USA, in 2017. He is currently a postdoctoral appointee at Argonne National Laboratory, Illinois, USA. His research interests include optimization and control theories with applications to power systems.